\documentclass[aps,prd]{revtex4}

\usepackage{amsmath, amsthm, amssymb, mathrsfs}
\usepackage{bm}
\usepackage{verbatim} % for commenting large sections

\begin{document}

\title{Background Independent Quantum Mechanics, Gravity,\\
 and Physics at Short Distance: Some Insights}
\author{Aalok} 
\email{aalok@uniraj.ernet.in}

\affiliation{Department of Physics, University of Rajasthan, Jaipur 302004 India;\\
 and Jaipur Engineering College and Research Centre (JECRC) 303905 India}

\date{\today}
%\pacs{} %need to use showpacs in displayclass options
%\keywords{} %need to use showkeys in displayclass options

%--------------------------------------------

\begin{abstract}
In the present discussion Background Independent framework of Quantum Mechanics and its possible implications in the
studies of gravity and Physics at short distance are addressed. The expression of the metric of quantum state space 
$g_{\mu\nu}$ which is intrinsically a quantized quantity, is identified in terms of Compton wavelength as:
 $[\langle\partial_{\mu}\psi\mid\partial_{\nu}\psi\rangle-
\langle\partial_{\mu}\psi\mid\psi\rangle\langle\psi\mid\partial_{\nu}\psi\rangle]
=\frac{1}{\lambdabar_{C}^{2}}(=\frac{m_{0}^{2}c^{2}}{\hbar^{2}})$. The discussion also sheds light on the notion of
neighborhood in quantum evolution.  
\end{abstract}

\maketitle

PACS number(s): 04.60.-m, 11.25.Yb\\

There is a prevailing feeling that either Quantum Mechanics (QM) or General Relativity (GR) or both should pave way for
new  geometrical feature in QM [1-4]. And an intensive follow up of this call for the extension of standard geometric
quantum mechanics [1-8] would be academically rewarding [1]. Physicists studying gravity have also shown considerable 
interest in the  geometric structures in QM in general and projective Hilbert space in specific [1-4, 9 ,10]]. However,
we feel that there is  enough information hidden in the standard geometric QM that is yet to be explored.
The present discussion aims to address Background Independent framework of Quantum Mechanics and its possible 
applications in the studies of gravity and Physics at short distance.\\
To begin with, we briefly discuss the basic tenets of standard geometric QM [1-4, 6, 7], and the background independent
 settings in which investigations are going on, to make it relevant to studies of gravity. Pure states are points
 of an infinite dimensional K$\ddot{a}$hler manifold on $\mathscr{P(H)}$ the complex projective space of the Hilbert
 space $\mathscr{H}$. Equivalently $\mathscr{P(H)}$ is a manifold with an almost complex structure. The
probabilistic interpretation lies in the definition of geodesic length on the space of quantum states (events).\\
The space of quantum mechanics (events) becomes dynamical and that the dynamical geometrical information is described in
 terms of a non-linear diffeomorphism invariant theory in such a way that the space of quantum events is non-linearly
 inter-related with the Hamiltonian- the generator of quantum dynamics. The distance on the projective Hilbert space is
 defined in terms of metric, called the metric of the ray space [1-4, 6-10] or the projective Hilbert space $\mathscr{P}$,
is given by the following expression in Dirac's notation:
\begin{equation}
ds^{2}=[\langle d\psi\mid d\psi\rangle-\langle d\psi\mid\psi\rangle\langle\psi\mid d\psi\rangle]
\end{equation}
This can be an alternative definition of the Fubini-Study (FS) metric, valid for an infinite dimensional
$\mathscr{H}$.\\
The metric in the ray space being treated by physicists as the background independent and space-time independent 
structure, can play an important role in the construction of a potential ''theory of quantum gravity''. The demand
 of background independence in quantum theory of gravity calls for an extension of standrd gemetric quantum
 mechanics [1-4]. It is an important insight which can be springboard for our proposed background independent
 generalization of standard quantum mechanics. For a generalized coherent state, the FS metric reduces to the metric on
 the corresponding group manifold [2]. Thus, in the wake of ongoing work in the field of quantum geometric formulation,
the work in the present discussion may prove to be very useful. The probabilistic (statistical) interpretation of QM
is hidden in the metric properties of $\mathscr{P(H)}$. The unitary time evolution is also in a way related to the
metrical structure [1, 2, 6-10] with Schr$\ddot{o}$dinger's equation in the guise of a geodesic equation on $CP(N)$. The
time parameter of the evolution equation can be related to the quantum metric \textit{via}:          
 \begin{equation}
(\Delta E)^{2}\equiv\langle\psi\mid H^2\mid\psi\rangle-\langle\psi\mid H\mid\psi\rangle^2;
\end{equation}
with $\hbar ds=\Delta Edt$.\\
And the Schr$\ddot{o}$dinger equation can be viewed as a geodesic equation on $CP(N)=\frac{U(N+1)}{U(N)\times U(1)}$ as:
\begin{equation}
\frac{du^a}{ds}+\Gamma_{bc}^{a}u^{b}u^{c}=\frac{1}{2\Delta E}Tr(HF_{b}^{a})u^{b}.
\end{equation}
Here $u^{a}=\frac{dz^{a}}{ds}$ where $z^{a}$ denote the complex coordinates on $CP(N)$, $\Gamma_{bc}^{a}$ is the
connection obtained from the Fubini-Study metric, and $F_{ab}$ is the canonical curvature 2-form valued in the holonomy
gague group $U(N)\times U(1)$. Here, Hilbert space is $N+1$ dimensional and the projective Hilbert space has dimenssions
$N$.\\
If the metric of quantum states is defined with the complex coordinates in the quantum state space, known as Fubini-
Study metric, it lies on the K$\ddot{a}$hler manifold or $CP(N)$, which is identified with the quotient set 
$\frac{U(N+1)}{U(N)\times U(1)}$.\\
The symmetries described by this quotient set have limitations. However, the most appropriate representation that seems to
satisfy the almost complex structure criteria is the Grassmannian. By the correspondence principle, the generalized 
quantum geometry must locally recover the canonical quantum theory encapsulated in $\mathscr{P(H)}$, also with mutually
compatible metric and symplectic structure, allows the framework for the dynamical extension of the canonical quantum
theory.\\
The Grassmannian:
\begin{equation}
Gr(C^{N+1})=\frac{Diff(C^{N+1})}{Diff(C^{N+1}, C^{N}\times {0})},
\end{equation}
 in the limit $N\to \infty$ satisfies the necessary conditions [10]. The Grassmannian is gauged version of complex
projective space, which is the geometric realization of quantum mechanics.
The utility of this formalism is that gravity embeds into quantum mechanics with the requirement that the kinematical
 structure must remain compatible with the generalized dynamical structure under deformation [10]. The quantum
 symplectic and  metric structure, and therefore the almost complex structure, are themselves fully dynamical. Time
 the evolution parameter in the generalized Schr$\ddot{o}$dinger equation is yet not deemed to be global and is thus
transformed in terms of  the invariant distance. The basic point as threshold of the BIQM is to notice that the evolution
equation (the generalized Schr$\ddot{o}$dinger equation) as a geodesic equation can be derived from an Einstein-like
equation with  the energy-momentum tensor determined by the holonomic non-abelian field strength $F_{ab}$ of the 
$Diff(\infty-1, C)\times Diff(1, C)$ type and the interpretation of the Hamiltonian as a charge. Such an extrapolation
 is logical, since $CP(N)$ is an Einstein space, and its metric obeys Einstein's equation with a positive cosmological
 constant given by:
\begin{equation}
R_{ab}-\frac{1}{2}Rg_{ab}-\Lambda g_{ab}=0.
\end{equation}
The diffeomorphism invariance of the new phase space suggests the following dynamical scheme for the BIQM as:
\begin{equation}
R_{ab}-\frac{1}{2}Rg_{ab}-\Lambda g_{ab}=T_{ab}.
\end{equation}
 Furthermore, 
\begin{equation}
\nabla_{a}F^{ab}=\frac{1}{2\Delta E}Hu^{b}.
\end{equation}
The last two equations imply \textit{via} Bianchi identity, a conserved energy-momentum tensor
\begin{equation}
\nabla_{a}T^{ab}=0.
\end{equation}
This taken together with the conserved ``current'' as 
\begin{equation}
j^{b}=\frac{1}{2\Delta E}Hu^{b};
\end{equation}
implies the generalized geodesic Schr$\ddot{o}$dinger equation. Thus equations (8) and (9), being a closed system of
 equations for the metric and symplectic structure do not depend on the Hamiltonian, which is the case in ordinary 
QM too. Moreover, the requirement of diffeomorphism invariance places stringent constraints on the quantum geometry.
 We have to have an almost complex structure for the generalized space of quantum events. This
 extended framework readily implies that the wave-functions labeling the relevant space are themselves irrelevant.
They are as meaningless as coordinates in General Relativity.\\
The metric of the quantum state space has been identified as background independent (BI) metric structure [1-4]. 
By appearance itself the invariance of the significance of geometric structure in equation (1) is apparent.
The reformulation of the geometric QM in this background independent settings gives us
many a new insights. Quantum states being unobservable and also due to $Diff(\infty, C)$ symmetry, make no sense 
physically, only quantum events do. This is quantum counterpart of the corresponding statement of the meaning
of space-time events in GR. Probability is generalized and is given by the notation of diffeomorphism invariant distance
in the space of quantum configurations.\\
As discussed repeatedly, the expression in equation (1) is the metric in the BIQM framework that leads to yet another
question: what this invariant and constant quantity stands for? The answer is revealed by the Klein-Gordon evolution.
And we emphasize that there are interesting facts associated with this geometry of quantum state space that cannot be 
ignored.  
The metric of the quantum state space, which is intrinsically a quantized quantity as\\
$g_{\mu\nu}=[\langle\partial_{\mu}\psi\mid\partial_{\nu}\psi\rangle-\langle\partial_{\mu}\psi\mid\psi\rangle\langle\psi\mid\partial_{\nu}\psi\rangle]$ had originally been derived from the expression 
$[\langle\psi\mid H^{2}\mid\psi\rangle-\langle\psi\mid H\mid\psi\rangle^{2}]$. 
As already shown [1], if we consider the relativistic evolution of quantum states by Klein-Gordon equation, it reveals
the reasons that give rise to the this covariant and invariant quantity as:
\begin{equation}
-\psi^{*}\nabla^{\mu}\nabla_{\mu}\psi=\frac{m_{0}^{2}c^{2}}{\hbar^{2}}\psi^{*}\psi.
\end{equation}

By using the definition of covariant derivative of the quantum state space in Dirac's notation, left hand side of
equation (10) could be rewritten in generalized manner [1] and the expression of the metric of quantum state space
is obtained as follow:   

\begin{equation}
\frac{1}{2}[(\psi^{*}\nabla_{\mu}\nabla_{\nu}\psi)+(\psi^{*}\nabla_{\mu}\nabla_{\nu}\psi)^{*}]\\
=-[\langle\partial_{\mu}\psi
\mid\partial_{\nu}\psi\rangle-\langle\partial_{\mu}\psi\mid\psi\rangle\langle\psi\mid\partial_{\nu}\psi\rangle].
\end{equation}
From which we find an interesting result:
\begin{equation}
[\langle\partial_{\mu}\psi\mid\partial_{\nu}\psi\rangle-\langle\partial_{\mu}\psi\mid\psi\rangle\langle\psi\mid\partial_{\nu}\psi\rangle]=\frac{m_{0}^{2}c^{2}}{\hbar^{2}}.
\end{equation}
The quantity in the right hand side is familiar one. It can be defined as square of inverse of the Compton's wavelength
as:
\begin{equation}
(\frac{m_{0}c}{\hbar})^{2}=\frac{1}{\lambdabar_{C}^{2}}.
\end{equation}
Thus one can think of the invariant $ds^{2}$ in the ray space evolving as multiple of inverse of the Compton wavelength.
As this final expression is valid irrespective of the choice of quantum states, we can draw this inference for all
quantum states in generality. 
At long wavelengths, once we map the configuration space to space-time, we have General Relativity. Turning off dynamics 
the quantum configuration space recovers the canonical quantum mechanics [10].\\
An equally important clue one comprehends from equation (13) that gives rise to the question, whether the presence
of a quantity such as inverse of distance squared imply the signatures of gravity or a cosmological constant in this
geometric structure? This is subject of rigorous investigations.\\
 Interestingly, the Compton wavelength at Planck scale:
\begin{equation}
(\lambdabar_{C})_{Planck Scale}=\frac{\hbar}{m_{Pl}c}=1.6\times 10^{-33} cms,
\end{equation}
is precisely the Planck's length. Thus, the lowest value that the Compton wavelength ceases to be, is the Planck's length
only.\\
We know that cosmological constant is the variance in the vacuum energy about zero mean. The variance $\Delta E$ as it 
appeared in one of the original propositions [6] of the metric of quantum states 
\begin{equation}
ds^{2}=\frac{(\Delta E)^{2}}{\hbar^{2}}dt^{2},
\end{equation}
leads to a natural question: what this uncertainty of energy stands for? Also, the conclusion of equation (13) is obvious
if the variance in energy $\Delta E$ in equation (15) could assume a typical value $(\Delta E)^{2}\sim (m_{0}c^{2})^{2}$. 
If the quantum state under consideration is the state of vacuum then it could be the variance in the vacuum energy as: 
\begin{equation}
(\Delta E)^{2}=\langle 0\mid H^{2}\mid 0\rangle-\langle 0\mid H\mid 0\rangle^{2}.
\end{equation}

 It is interesting to note that there is something physical in the right hand side of equation (15) which appears as a
geometrical form in the left hand side of the equation. The invariant $ds$ in the metric structure of quantum states
is not distance in the dimensional sense, it is neighborhood in the topological sense. It is the infinitesimally
small neighborhood implied by this expression which fills the space. This expression of metric of quantum states as it
appeared in one of its original propositions [6] was later generalized in the quantum state space.
As suggested by T. W. Kibble [5] in the context of proposed generalization of quantum mechanics that the states that are
in a sense defined near vacuum can be represented by vectors in the tangent space $T_{\nu}$, and that on $T_{\nu}$ one has
all the usual structure of linear quantum mechanics expressed in the local coordinates. However, we need to specify what
is meant by ''nearness'' to the vacuum. At each point on the space-time manifold, the space is locally flat. Locally, the
vacuum energy is fixed by the quantum theory in the tangent space, which is also the case in the Matrix theory [10]. 
Gauging QM generically breaks Super-Symmetry. We do not have globally defined super-charges in space-time in the
correspondence limit. This also explains- why there is cosmological constant [10].
The detail study will appear elsewhere.\\

%-------------------------------------------------------
\begin{acknowledgments}

The author wishes to thank Prof. A. Ashtekar for explaining the need and importance of background independent
quantum mechanics.
\end{acknowledgments}

\end{document}